\documentclass[]{scrartcl}

\usepackage[utf8]{inputenc}
\usepackage[english]{babel}
\usepackage[hidelinks]{hyperref}
\usepackage{graphicx}
\usepackage{listings}
\usepackage{natbib}
\usepackage{url}
\usepackage{wasysym}

\renewcommand\cite[1]{\citep{#1}}

\title{Feature Interactions on Steroids: \\On the Composition of ML
Models}
\author{Christian K\"astner, Eunsuk Kang, Sven Apel}
\begin{document}
\maketitle

Much has been written about how machine learning changes software
engineering practices in big and small ways~\citep[e.g.,][]{sculley2015hidden,amershi2019software,ozkaya2020really,KK:ICSESEET20}. Among all differences, the
lack of specifications comes up again and again and has rippling effects
on all kinds of other practices. Here, we discuss how it drastically
impacts how we think about divide-and-conquer approaches to system
design, and how it impacts reuse, testing and debugging activities.

Traditionally, specifications provide a cornerstone for compositional
reasoning and for the divide-and-conquer strategy of how we build large
and complex systems from components, but those are hard to come by for
machine-learned components. While the lack of specification seems like a
fundamental new problem at first sight, in fact software engineers
routinely deal with iffy specifications in practice: we face weak
specifications, wrong specifications, and \emph{unanticipated
interactions} among components and their specifications. Machine
learning may push us further, but the problems are not fundamentally
new. Rethinking machine-learning model composition from the perspective
of the \emph{feature interaction problem}, we may even teach us a thing
or two on how to move forward.

\hypertarget{challenges-in-composing-ml-models}{%
\section{Challenges in Composing ML
Models}\label{challenges-in-composing-ml-models}}

\begin{figure}[ht]
\includegraphics[width=\linewidth]{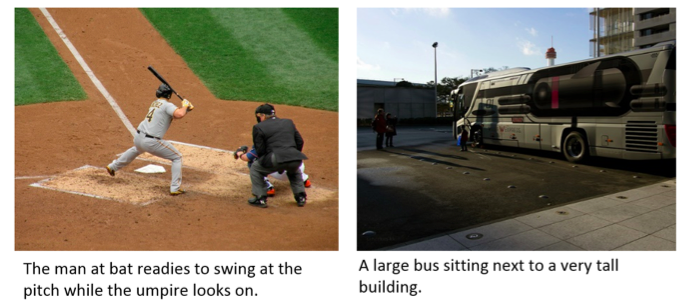}
\caption{Example images and corresponding captions from
the COCO competition~\cite{coco}}
\label{fig:example}
\end{figure}

Let's jump right into the problem with an example from the excellent
AAAI'17 paper \emph{``On
human intellect and machine failures: Troubleshooting integrative
machine learning systems}''~\cite{nushi2017human}. The paper discusses a system with the goal
to generate captions for images, which was part of a
machine learning
competition (see Fig.~\ref{fig:example}).

Notice that \emph{the problem and what makes a good solution is somewhat
fuzzy}, which will become a key point in a second: Given an image,
provide a description for that image. A number of examples of good
descriptions are provided (for training and evaluation), but there is no
clear, formal specification of what makes a caption \emph{``correct''}
or \emph{``good.''} It is important to note that we use machine learning
(in this case) exactly because we do not have a specification in the
first place! The goal is to automatically find heuristics that best
imitate human (or any other) behavior without explicitly specifying
it---this is \emph{inductive reasoning}: learning rules from
observations. In fact, in said research competition, the quality of a
solution is eventually judged by humans, who rate the quality of
captions generated for example images, not by checking results against
fixed rules. Needless to say, different judges may come to different
conclusions.

Building such a captioning system with a single model is a huge step.
Instead, the state-of-the-art approach discussed (which
comes from another paper~\cite{fang2015captions})
\emph{decomposes the problem} into three steps, each with a separate ML
model, as shown in Figure~\ref{fig:caption-arch}:

\begin{figure}[ht]
\centering
\includegraphics[width=\linewidth]{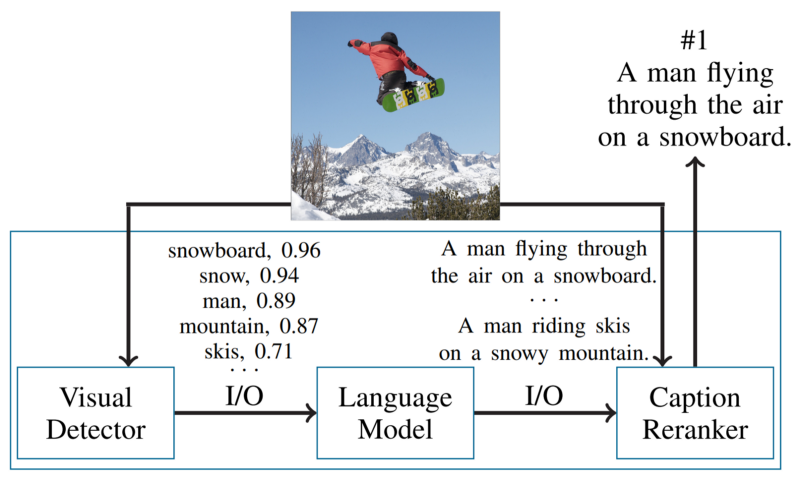}
\caption{Three-step architecture of the image captioning system \cite{nushi2017human}}
\label{fig:caption-arch}
\end{figure}

\begin{itemize}
\item
  \textbf{Visual detector:} First, a convolutional neural network is
  trained as a visual detector with a vocabulary of 1000 objects, which
  attempts to identify what kinds of objects are present in the image
  and produces corresponding confidence scores. This component takes an
  image and produces scores for different objects: \lstinline`Image -> List[(Objectname, ConfidenceScore)]`.
\item
  \textbf{Language model:} Second, a maximum-entropy language model is
  trained on many image captions and used to generate plausible
  sentences with the words from the detected objects. This component
  will generate 500 sentences with different subsets of the detected
  objects. The key here is to generate sentences that contain some of
  the words for the detected objects and that are likely and plausible
  based on typical grammatical patterns and word frequencies appearing
  generally in image captions---but without any actual information
  about the image beyond what objects are (possibly) in it. For example,
  the model would hopefully generate ``a mountain covered in snow'' as a
  more likely phrase than ``a mountain skiing downhill''. For each
  sentence, the model also produces a likelihood score that indicates
  how well this sentence mirrors typical image captions:
  \lstinline`List[(Objectname, ConfidenceScore)] -> List[(Sentence, Likelihood)]`.
\item
  \textbf{Caption ranker:} Finally, a deep multimodal similarity model,
  consisting of two neural networks, takes both the original image and
  the generated sentences, extracts features from both, and scores the
  combination (don't ask for details, please). The component then picks
  the sentence with the best score. Internally, the ranker uses multiple
  features, including objects detected in the image (similar to the
  visual detector) and the similarity between the vector representations
  of images and captions. This component has the signature: \lstinline`Image -> List[(Sentence, Likelihood)] -> Sentence`.
\end{itemize}

At a first glance, this looks like a great divide-and-conquer story: We
can think about how a human would break down the task and then let the
solution (partially) mirror some of those steps, such as recognizing
objects and formulating short plausible sentences about those objects.
Each component can be developed, tested, and even deployed
independently, using different modeling and implementation techniques.
We can describe the interface (signature, types, invariants, etc.),
deploy them as microservices, and roughly describe the intended behavior
of each component. These components (especially, the detector and the
language model) may be general-purpose and reused in other contexts.
Moreover, independent teams can work on improving each of the
components. At a technical level, it is fairly straightforward to
compose a system with these three components.

\hypertarget{problem-1-assigning-blame}{%
\subsection{Problem \#1: Assigning
blame}\label{problem-1-assigning-blame}}

At a second glance, various problems emerge: If something goes wrong
somewhere, and we get a poor caption for an image, what caused the
problem? How can we improve the system? How do changes to one component
affect the rest of the system? For example, in the AAAI paper, the
system is reported to create the caption \emph{``A blender sitting on top of a
cake''} for the image in Figure~\ref{fig:caption-problems}, but why?

When looking at the individual steps in Figure~\ref{fig:caption-problems}, the authors see that potentially
all components can be blamed:

\begin{figure}[ht]
\centering
\includegraphics[width=\linewidth]{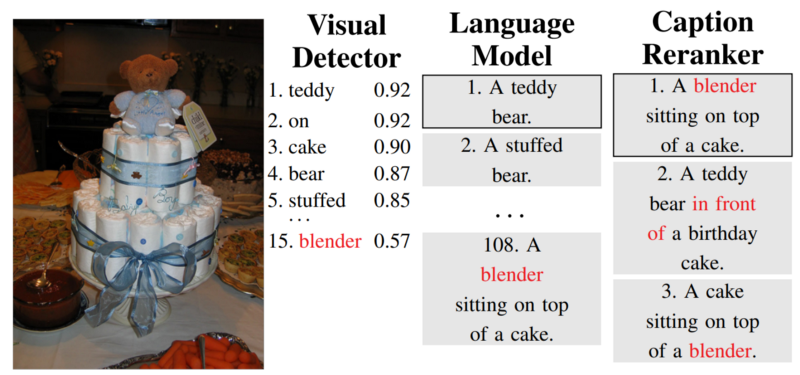}
\caption{Problems in each component leading the to the poor caption \emph{``A blender sitting on top of a
cake''}
 (from \citet{nushi2017human})}
\label{fig:caption-problems}
\end{figure}

\begin{itemize}
\item
  First, the visual detector detects a blender where there is none. To
  its credit, the confidence score is fairly low, but we can clearly
  consider this a mistake.
\item
  Second, the language model takes these words and suggests ``a blender
  sitting on top of a cake'' as a plausible sentence containing the
  detected words ``blender'', ``on'', and ``cake''. We might argue to
  give this component some slack, because ``blender'' probably should
  not be part of its input, but even for the given input words, the
  produced sentence does not seem quite right, given that a ``sitting
  blender'' is not a very plausible expression, i.e., the model shows
  low common-sense awareness. It's ranked 108 out of 500 generated
  sentences.
\item
  Finally, the ranking algorithm picks that sentence, even though it
  includes a word with a low object detection score. Actually, none of
  the top-scored sentences seems particularly great.
\end{itemize}

So, no component is behaving perfectly or compensating enough for
problems in other components. It seems hard to blame a single component.
There are no clear responsibilities or boundaries between the
components. If we had only a limited budget for improving the system,
which component should we start with?

\hypertarget{problem-2-nonmonotonic-errors}{%
\subsection{Problem \#2: Nonmonotonic
errors}\label{problem-2-nonmonotonic-errors}}

\begin{figure}[ht]
\centering
\includegraphics[width=\linewidth]{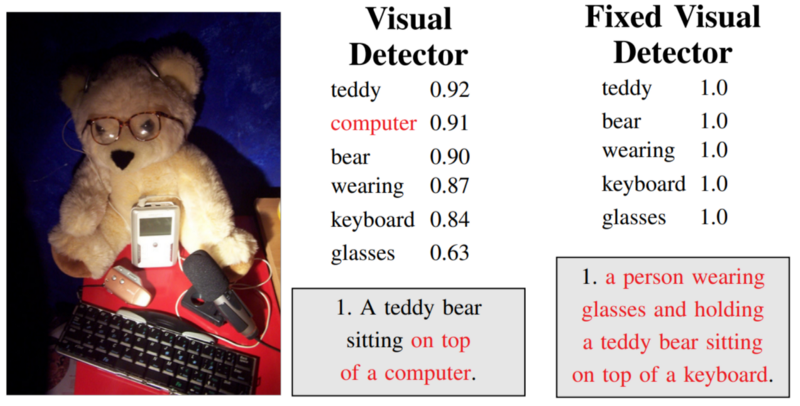}
\caption{Results get worse if output of first component is fixed
(from \citet{nushi2017human})}
\label{fig:caption-fix}
\end{figure}

It gets worse though. One would hope that by improving any one component
of the system, overall results would improve. Unfortunately, this is not
always the case. In Figure~\ref{fig:caption-fix}, we show another example from the paper:
The initially generated caption \emph{``A teddy bear sitting on top of a
computer''} is not great, given that there is no computer in the
picture. However, if we fixed the object detector for this example
(i.e., removing ``computer'' and setting confidence of all objects
actually present to 1.0), the caption get's clearly worse: \emph{``a
person wearing glasses holding a teddy bear sitting on top of a
keyboard.''} It's hard to give a causal explanation for what happens
here, but it seems that, with larger emphasis on ``glasses'', a sentence
that includes glasses is picked, but the language model probably rates a
``teddy wearing glasses'' as less plausible than ``a person wearing
glasses'' and thus makes up an object not originally detected.

Note that the language model was somewhat robust to imperfect inputs:
The component and the overall system usually works even with mistakes in
the object detector. The language model does not expect or require
perfectly accurate inputs and making up new objects for more plausible
captions is perfectly okay. Generally, some models may even learn to
compensate for common mistakes in their inputs.

\begin{figure}[ht]
\centering
\includegraphics[width=\linewidth]{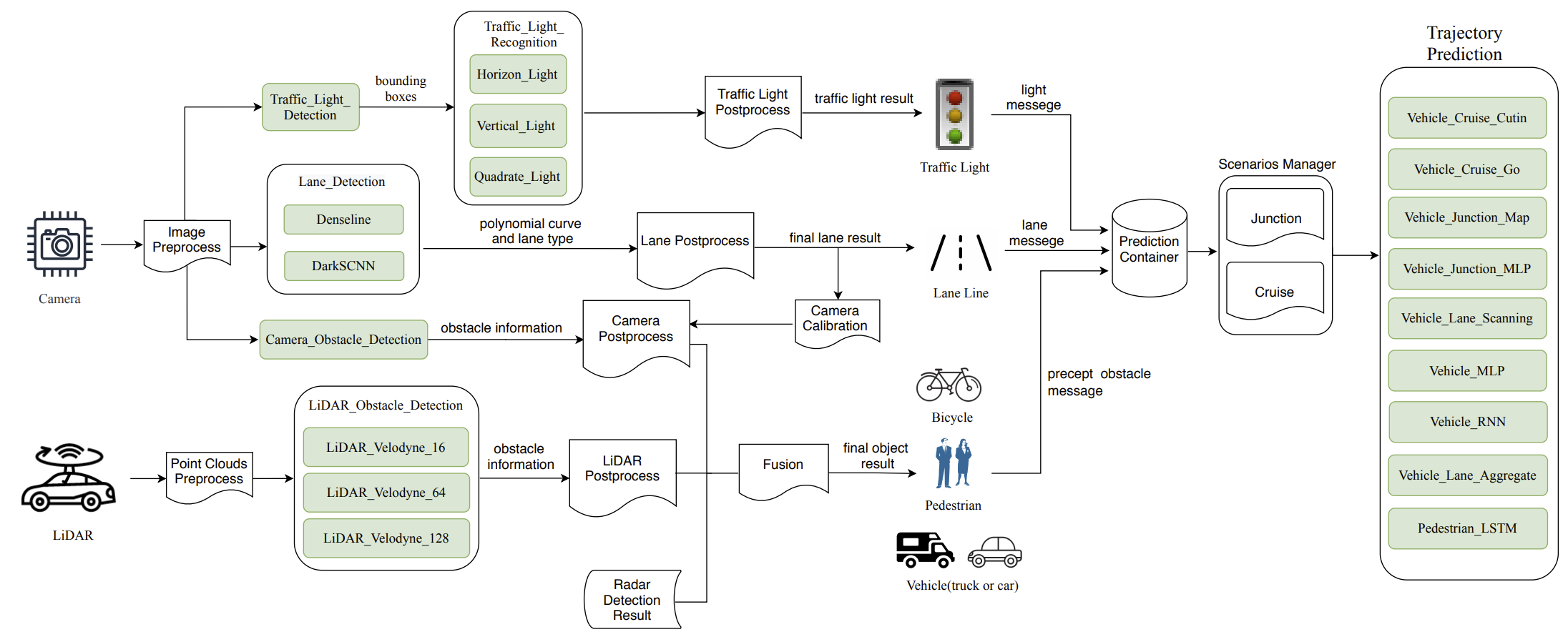}
\caption{High-level architecture of Apollo and the 18 ML models composed in
a system to autonomously drive a car---graphic from \citet{peng2020first}; see the paper for details on the models and their
inputs and outputs.}
\label{fig:apollo}
\end{figure}

The key problem is that it is not always the case that improving one
component improves the overall system's quality. So, if we already have
a problem with composing three models, how are we expecting to build
reliable systems with many more ML and non-ML components, say the 18
models in Baidu's self-driving car system Apollo shown in Figure~\ref{fig:apollo}.

\hypertarget{the-root-of-the-problem-lack-of-specifications}{%
\section{The Root of the Problem: Lack of
Specifications}\label{the-root-of-the-problem-lack-of-specifications}}

The problem with reasoning about and composing ML models is that we do
not have clear specifications for what they are supposed to do. A
traditional specification would tell us for a given input whether a
produced output is correct or not. Specifications are powerful for
divide-and-conquer design strategies, for modular reasoning, and for
assigning blame.

\begin{figure}[ht]
\centering
\includegraphics[width=.6\linewidth]{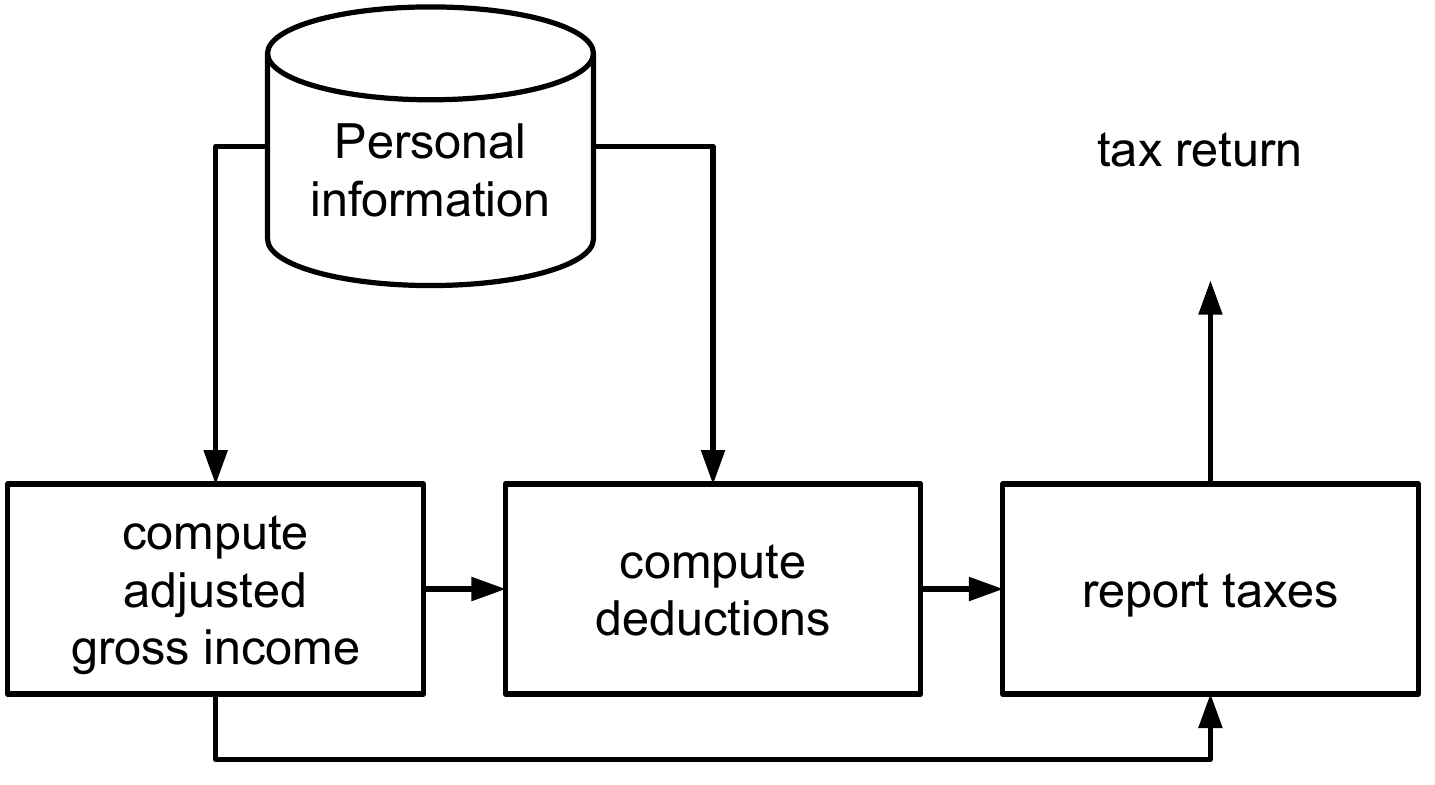}
\caption{Possible system decomposition for computing taxes}
\label{fig:tax}
\end{figure}

Let us contrast the ML captioning system with the most boring non-ML
system we can think of: \emph{computing taxes}. Also here, we probably
do not want to do all computations in a single step, but break the
system down into components. For symmetry let us just stick to three
components again, illustrated in Figure~\ref{fig:tax}: one for computing the adjusted gross income (some
notion of what income counts for tax purposes, do not ask), one for
computing deductions (where some deductions depend on the income
brackets), and one module to generate the text of the actual tax return.

We can again define an interface for each component (\lstinline`computeAGI: Data -> Int`; \lstinline`compute-deductions: Data -> Int -> Int`; \lstinline`report: Int -> Int -> String`), but more importantly, we know exactly what the system and each
component is supposed to do, what its responsibilities are, and what it
expects from other components.

Based on provided inputs, a computed tax return is \emph{either correct
or wrong}. It is not ``pretty good'' or ``95\% accurate'' as judged by a
human expert; it either works as specified and produces the right result
or it does not. We also would not accept it producing correct tax
returns for 98\% of all users, but would instead consider it as a
problem if it produced a wrong result for any single valid input.

If the output is wrong, we can \emph{assign blame} by checking whether
the income, the deductions, and the report were computed correctly based
on their respective given inputs. Again, each component's output is
either correct or wrong for the provided inputs. If a person's reported
deductions are wrong, but that's due to incorrectly computed adjusted
gross income, we would not blame the deductions or reporting component
(they work correctly for the inputs they were given) but would blame the
compute-adjusted-gross-income component that produced the wrong value
used by the others.

Similarly, if we fix a bug in a component to produce the correct
component output, this should never decrease the correctness of the
system. We may expose further bugs in other components that previously
were suppressed by the bug that is now fixed upstream, but those other
bugs were already there.

Decomposition works often in traditional software systems because we can
specify the behavior of individual components. With \emph{deductive
reasoning} based on logic, we can understand the consequences from
combining two modules in terms of the composed specifications. This is a
cornerstone to a sound divide-and-conquer strategy and
\emph{modularity}. If each component produces the correct outputs for
the given inputs, so does the composed system---this is, at least, the
idea. If something breaks, we analyze which component is responsible.
Specifications even open the door for \emph{modular reasoning} and
\emph{information hiding}, where to understand a component, one only
needs to understand its inputs and the interfaces and specifications
(but not internal implementations!) of all other used components. This
is how software engineers were able to build and compose so much
reusable code in libraries and software ecosystems, where we can compose
layers and layers of abstractions, but this is a discussion for another
day.

\textbf{Real Specifications\ldots{}} To be frank, we should acknowledge
that, in practical software engineering, we almost never see the strong
formal specifications as covered in computer-science textbooks. Most
specifications are textual and informal, such as API documentation for
most libraries. We also routinely work with systems that have only weak
or implicit specifications that cover only part of the behavior, miss
side effects, or do not cover important qualities like resource
consumption---if we have any written specifications at all. The fact
that most specifications are informal or not even written down at all
does not prevent us from reasoning about compositions: We still can
\emph{in principle} provide a specification on demand, and we can use
that understanding of the specification to assign blame (or renegotiate
with the component's consumer what the specification should be).

\hypertarget{back-to-machine-learning}{%
\section{Back to Machine
Learning\ldots{}}\label{back-to-machine-learning}}

For machine-learned components, we do not have any meaningful
specifications in the traditional sense that could tell us whether an
output is correct for any given input which we could use to judge
whether a model works correctly for all possible inputs. For many
problems, we already have a hard time capturing what it means for a
single prediction of a model to be ``correct'' (multiple humans may not
agree).

Even if we had a strict binary notion of correctness for any single
prediction, we would \emph{not} expect a model to make correct
predictions for every possible input. In traditional software
engineering, if we find an input-output pair that does not match the
specification, we consider this as a \emph{bug}. In contrast, in machine
learning, we know to expect occasional wrong predictions. For some hard
problems, we would actually be quite happy if even just 30\% of all
answers were correct, for other problems 99\% accuracy is okay but not
great. In line with the aphorism \emph{``\href{https://en.wikipedia.org/wiki/All_models_are_wrong}{all models are wrong, but some
are useful},''} we do not evaluate the \emph{correctness} of a model but
whether the model mostly \emph{fits} our problem. We evaluate fit as a
relative measure, typically some form of \emph{prediction accuracy} over
some (hopefully representative) input-output example, not as a binary
correctness criteria for the model. This does not compose nicely.

While they do not have specifications in the traditional sense that
relate any inputs and outputs, there are still some things we know and
can reason about. We still have an intuition or goal for what a model is
supposed to do, be it detecting objects or generating plausible
sentences. We may have some invariants, such as fairness requirements
that we can even state formally. And we usually also have an intuition
of how we expect components to behave when composed, as in our image
captioning system, even though we may not be able to logically deduce
that behavior from the component's behavior. That is, practical non-ML
software and machine-learned models are not at opposing extreme ends on
a spectrum between strong and no specifications and our activities in
developing and assuring them should reflect this. While there are
important differences, there are also lots of similarities, as we will
see.

\emph{Side note:} In the broader field of artificial intelligence, there are
numerous techniques that go beyond learning heuristics from
observations. Beyond machine learning, \emph{symbolic} AI techniques are
often based on deductive reasoning~\cite{russell2020artificialintelligence}, often with nice compositional
properties, for example, many traditional planning and control problems
encoded as constraint satisfaction problems. Also probabilistic
reasoning about uncertainty is compatible with specifications and
facilitates clear reasoning about composition, though researchers still
actively work on technical challenges to scale technical inference or
verification steps needed for many problems~\cite{Pfeffer2016-ii}.

\hypertarget{the-limits-of-decomposition-feature-interactions-emergent-behavior}{%
\section{The Limits of Decomposition: Feature Interactions
(Emergent
Behavior)}\label{the-limits-of-decomposition-feature-interactions-emergent-behavior}}

While it may seem that decomposition in traditional software systems is
obvious and clean, unfortunately, that is not always the case either. It
is not even just a question of whether we write formal or informal,
strong or weak specifications, but a question of whether this would be
possible in the first place. The study of the \emph{feature interaction
problem} in software engineering has made this ample clear.

So what are feature interactions actually? Sometimes a mandated
decomposition of a system is simply way too optimistic, and we are
missing interactions that should have been obvious from a component's
specifications, but were not. In most of these cases, we make,
accidentally or even intentionally, wrong assumptions about the
environment and about how a component interacts with the environment and
other components (see environmental assumptions
in \emph{the
World vs the Machine}~\cite{jackson1995world}). We say that
two components (or features) \emph{interact} in unanticipated ways,
giving rise to a \emph{feature interaction} or say that new behavior
\emph{emerges} from the composition of multiple components, referred to
as \emph{emergent properties} in systems engineering. Note ``feature
interactions'' here is not to be confused with interpreting
how data
features interact within an ML model~\cite{molnar2020interpretable}.

There are many examples of feature interactions, and they typically
follow a common pattern: We decompose the system into components, then
design, develop and test these components separately, but finally
observe unexpected behavior when composing them. The canonical example
is a \emph{call forwarding} feature in a phone system that competes with
a separately developed \emph{call waiting} feature on how to respond to
the same call on a busy line, but there are many examples where
components in cars or smart home systems interfere with each other when
combined. We can usually blame each problem on weak specifications that
did not anticipate the interaction, but in general the problem is much
bigger.

\hypertarget{decomposition-fallacies-in-complex-systems}{%
\subsection{Decomposition fallacies in complex
systems}\label{decomposition-fallacies-in-complex-systems}}

In complex systems, especially systems that live in the real world, it
is often not possible to nicely separate behavior and divide a problem
into subproblems cleanly. Behavior in many real systems is antimodular.

\begin{figure}
\centering
\includegraphics[width=.8\linewidth]{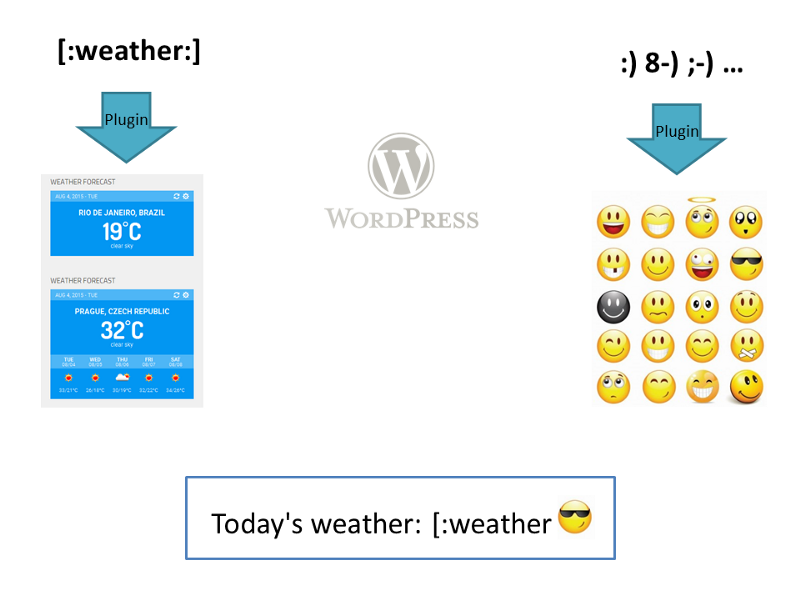}
\caption{Interaction between two independently developed and tested plugins in a blogging software}
\label{fig:wordpress}
\end{figure}

Let's postpone problems with the real world for a second and consider an
example of feature interactions at the software level: two WordPress
plugins that try to transform the same part of a blog post:
``\lstinline`[:weather:]`'' and ``\lstinline`:]`'' but, depending on their order, the
output may differ. The behavior of both plugins can seemingly be well
specified in isolation in that each plugin takes the blog post and
modifies it in an intended way, but we may not get the result we expect
if we compose them. Conceptually the problem is that the plugins do not
necessarily get the original blog post, but instead may receive input
already modified by other plugins, without being aware of these plugins,
possibly resulting in mangled output like ``\lstinline`[:weather`\smiley{}'' (see Fig.~\ref{fig:wordpress}).
The interaction is clearly foreseeable if we had known about the other
plugin and thought about it (overlapping preconditions, resolvable by
explicit ordering), but there are just so many plugins and potential
interactions to consider---after all, the number of potential
interactions among plugins grows exponentially with the number of
plugins. It is not obvious, though, how one could come up with a better
specification for each plugin, without thinking through all possible
plugin compositions upfront: If both plugins explicitly assumed the
original blog post as input, they would not be composable;
alternatively, if plugins always had to assume arbitrary other
transformations on their input first, they could make no meaningful
specifications about how their behavior relates to the user's blog post.
That is, we have to look at specific combinations of components and
cannot just reason about them separately, largely breaking our divide
and conquer strategy.

\begin{figure}
\centering
\includegraphics[width=.8\linewidth]{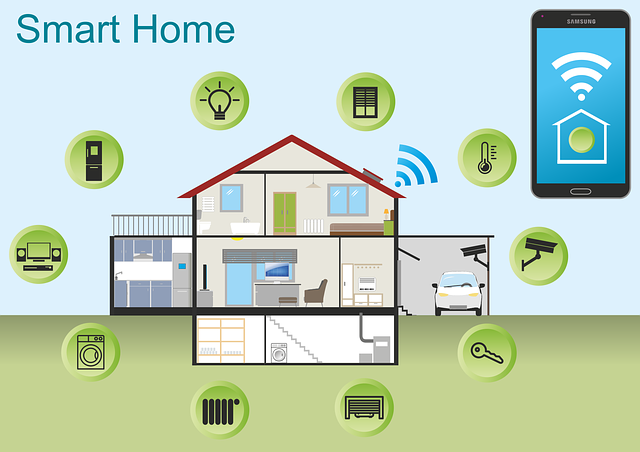}
\caption{Smart home system in which all kinds of independent devices can make decisions, possibly interacting in unintended ways.}
\label{fig:homeautomation}
\end{figure}

It gets much worse when the \emph{real world}™ is involved, as in
cyber-physical systems: Consider a home automation system with a heating
component, a ceiling fan component, and a component to control opening
windows. We could specify the behavior of each controller to some extent
and reason about how each component interacts with the environment.
However, in the environment, the actions of the different components may
influence each other through physical processes (e.g., warm air from the
heater moves through the ceiling fan to the open window). They may also
compete for the same resources, such as electricity or human attention.
To truly understand how the different components behave in concert in
the home automation system, we would need to fully understand the
environment, in addition to the actual components. For example, if we
wanted to fully specify how each component influences the room
temperature, such specification may need to involve aerodynamic physics
and the specific layout of the house and a clear understanding of how
other components behave---such a full environment model is clearly
unrealistic in practice. And even if we could model the environment, we
could not reason about components individually, but only about the
system as a whole. As requirements-engineering researcher Michael
Jackson framed it during a discussion at a Dagstuhl seminar: \emph{``the
physical world has no compositionality.''}

\hypertarget{decomposing-anyway}{%
\subsection{Decomposing anyway}\label{decomposing-anyway}}

A key insight from the study on feature interactions and systems
engineering broadly is that, even when systems are complex, we decompose
them anyway, and, necessarily, we make simplifying assumptions that may
not actually hold. Humans simply cannot deal with complexity beyond a
certain scale---we do not have the \emph{cognitive capacity}. We need
to abstract and decompose. We may make simplifications out of ignorance,
but we may also make them knowing very well that our decompositions are
not perfect. We may intentionally create specifications that are weak or
even wrong for specific cases. We may do this intentionally for good
reasons, hoping that we can resolve issues at composition time.

For example, in WordPress, we may pass a partially processed blog post
through multiple plugins knowing that we cannot cleanly separate their
effects. In the home automation setting, we separately design components
knowing that we are unlikely to ever understand how exactly they
interact in a specific environment. In fact, 
the \href{https://en.wikipedia.org/wiki/Feature_interaction_problem}{feature
interaction problem} gained popularity when, in the early 1990s,
telephone systems became more and more sophisticated and added more
features, such as call forwarding, that were deployed in different
hardware components of larger interconnected telephone networks with
different network providers, customer devices, and many device
manufacturers---complexity soon exceeded the cognitive capacity of
developers, and it became hard to foresee interactions within a single
device, let alone between networked devices from different
manufacturers. In all cases, a divide-and-conquer approach is
unavoidable to cope with complexity, even knowing that decompositions
are not clean. We simply have no other choice to deal with real-world
complexities in current engineering practices.

While humans are forced to decompose complex systems to handle
complexity, the good news is that, \emph{most of the time}, a
divide-and-conquer approach pays off and imperfect decompositions work.
Most of the time, we decompose systems, making a number of assumptions
that we know are not perfectly true, hope to catch inadvertent
interactions at composition time or at runtime, and mostly it works out
in practice. For example, most plugins work well together in complex
systems such as WordPress, Chrome, and Thunderbird. In the home
automation system, we can use control mechanisms and dynamically adjust
for interactions rather than to plan the exact interactions of
components. The problem are those remaining unanticipated interactions
from too weak or wrong specifications that surprise us, sometimes badly
with severe consequences, such as a fire control system malfunctioning.

\hypertarget{coping-with-feature-interactions}{%
\section{Coping with Feature
Interactions}\label{coping-with-feature-interactions}}

Feature interactions are not a new phenomenon. They have been actively
studied, at least, since the 90s' focus on telecommunication systems,
and as \emph{emergent properties} they are an important aspect of system
engineering. The community has learned how to build systems that work
reasonably well despite weak or even partially wrong component
specifications. Dedicated analysis, design, and control techniques
anticipate and compensate for these kinds of modularity problems.

While not a silver bullet, it is likely that there are insights from a
long tradition of coping with feature interactions that may help us to
better understand and build systems composed with multiple
machine-learned components and traditional components. In both worlds,
we explicitly deal with modularity and composition problems stemming
from weak or missing specifications. The thoughts and parallels below
are a starting point.

\hypertarget{detection-through-testing}{%
\subsection{Detection through
testing}\label{detection-through-testing}}

The first most obvious insight is that unit testing or component
verification are clearly not sufficient, even in traditional software
systems. That is, it cannot be sufficient to check whether each
component meets its specification, but we also need to test whether the
composed system as a whole indeed behaves as expected. This is where the
importance of \emph{integration testing} and \emph{system testing} comes
from.

\begin{figure}[ht]
\centering
\includegraphics[width=\linewidth]{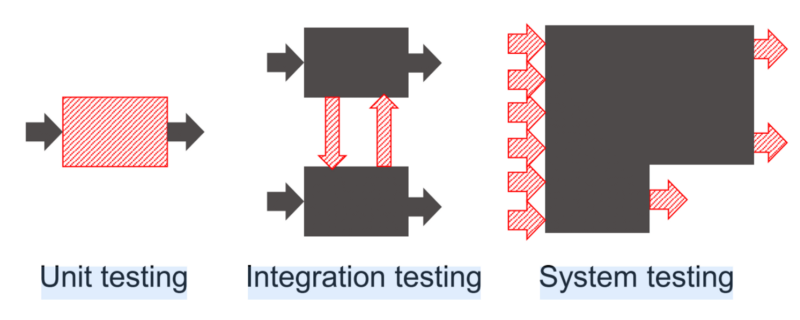}
\caption{Testing at different levels: Unit testing, integration testing, system testing}
\label{fig:testing}
\end{figure}

This observation holds also for systems with machine-learning
components, which is just another reminder not to stop at measuring
prediction accuracy on a test dataset (which itself can already be
challenging to do well), but to evaluate the entire system in production
and measure system goals (e.g., sales, customer satisfaction).

Considering feature interactions can also influence what kind of tests
to write for integration testing. Anticipating potential feature
interactions, we may want to write tests for behavior that should be
invariant whenever the component is included in a system. For example,
both plugins of the WordPress example could come with a test case
checking that the plugin correctly transforms a relevant blog post
independent of any other plugins installed (e.g., the Weather test would
fail if the Smiley plugin is active at the same time, thus detecting the
interaction)~\cite{NKN:ICSE14}:

\begin{lstlisting}[language=PHP]
function testWeather() {
   if (plugins.contains('weather')) {
       output = runWebPage('index.php', 'Today is [:weather:].');
       assertMatch(output.getElementByXPath('/html/body/div[1]'),
                   'Today is \D+ degree.');
   }
}
function testSmiley() {
   if (plugins.contains('smiley')) {
       output = runWebPage('index.php', 'Hey there :]');
       assertMatch(output.getElementByXPath('/html/body/div[1]'),
                   '.*<img "...">');
   }
}
\end{lstlisting}

In systems with machine-learned components, one can imagine similar
tests of contributions of individual components to the overall system
behavior, for example, the image captioning system should always produce
short descriptions independent of what objects are detected or how the
results are ranked. The traffic light detection in a self-driving car
should reliably detect obvious examples of traffic lights independent of
other components that may, say, involve detection under challenging
conditions involving a map or filter inputs against adversarial attacks.
Here, the test case could be represented with a test dataset on which to
achieve consistently high prediction accuracy, independent of other
components.

\hypertarget{detection-through-better-specifications-formal-specifications}{%
\subsection{Detection through better specifications, formal
specifications}\label{detection-through-better-specifications-formal-specifications}}

A long history of research on feature interactions has shown that better
requirements engineering can help to anticipate interactions, often
through systematic inspection of potential interaction points, or
actually writing down and mechanically checking compatibility of
specifications.

With strong specifications, we can detect conflicts between multiple
component specifications and the system specification. For example, we
could notice overlapping preconditions of the WordPress plugins'
transformations and thus detect a conflict at design time. If
specifications are written in a machine-readable format, various
analysis tools such as theorem provers and model checkers can help to
reason about potential composition problems.

Weak and informal specifications can still be useful to detect many
problems. For example, \emph{goal models} that describe the goal of each
component can help to identify overlapping and conflicting goals (e.g.,
heating, cooling, and opening windows in the home automation scenario).
Similarly \emph{resource models} that describe which external resources
components consume or control can help to detect conflicts, such as
multiple home automation components competing for electricity or human
attention in certain situations or a smart switch controlling the power
supply depended on by a space heater.

Even without tools and formal specifications or models, just by looking
at pairs of components, it is often possible to detect possible
interactions during a manual inspection step. Such a manual pairwise
analysis is quite common in practice for detecting possible interactions
(researchers have actually
held competitions
to find interactions within requirements in the 90s \cite{calder2000feature}), and it is often
guided by looking at probable interaction points, accessed resources,
environment assumptions, goals, and preconditions of components. In their survey paper, \citet{nhlabatsi2008feature} provide a concise overview of common kinds of conflicts and different
kinds of interactions that can help to guide an inspection. If a
possible interaction has been detected, it may be worth writing down
additional specifications about how to resolve it appropriately at
runtime (e.g., the weather plugin having priority over the smiley
plugin).

In a machine-learning context, we may be able to reason to some degree
about goals, resources, and maybe even weak specifications to detect
certain kinds of possible interactions, especially if models interact
through the environment and shared resources. However, given how hard it
is to provide even weak specifications, leveraging system design might
be a more promising solution.

\hypertarget{design-for-isolation-and-domain-specific-resolutions}{%
\subsection{Design for isolation and domain-specific
resolutions}\label{design-for-isolation-and-domain-specific-resolutions}}

Likely the most powerful strategy in addressing the feature interaction
problem is to prepare for unknown interactions as part of the design
process. That is, instead of trying to identify specific interactions
with testing or requirements analysis, we anticipate the possibility of
some unknown interactions as part of the system design: We design the
system to (1) prevent certain interactions by \emph{isolating}
components or (2) prepare for automatically \emph{resolving}
interactions when they eventually take place.

\paragraph{Isolation:} A common strategy to avoid interactions is to
isolate the effects of individual components. For example, in WordPress,
plugins do not transform the entire page, but individual parts, such as
the blog post or the header links, such that plugins changing different
parts do not interfere (though technical enforcement is not as strong as
it could be). In Android, apps are largely sandboxed and cannot access
each other's internal state, each other's files, or each other's screen
output.

\begin{figure}[ht]
\centering
\includegraphics[width=.4\linewidth]{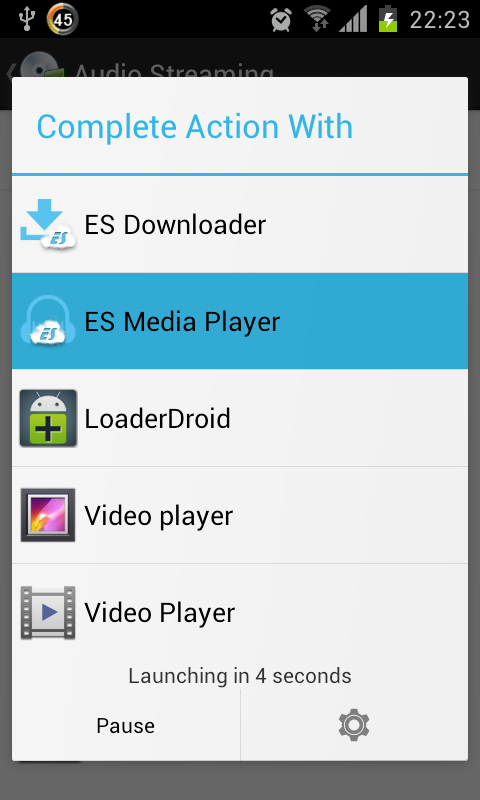}
\caption{System-wide interaction resolution in Android: When multiple apps can respond to a request, the system asks the users.}
\label{fig:android}
\end{figure}

Even when possible, \emph{full} isolation is rarely desired, as
components should often work together in intended ways. Such designs
will therefore typically allow specific kinds of interactions, but often
require to use permitted and possibly controlled communication channels.
For example, WordPress allows multiple plugins to extend individual
extension points, each allowing only limited transformations. Android
provides well-defined communication channels (``Intents'' and system
APIs) controlled by the operating system, which ensures that only
certain messages in certain formats can be exchanged; the system can
even intercept, modify, or block messages at runtime if it serves the
system specifications, such as not making phone calls without the
corresponding app permissions. This design limits interactions to much
more controllable points.

\textbf{Resolution:} An interaction can be resolved manually once it has
been detected by adding \emph{coordination logic.} Resolution typically
consists of one component overwriting the behavior of another (e.g.,
disable cooling while the heater is running), ordering components (e.g.,
always run the smiley plugin last), adding some logic combining results
(e.g., when multiple components suggest conflicting target temperatures,
take the average). Anticipating the need for resolution, e.g., by
enabling flexible reordering, will make it easier for developers or end
users to resolve interactions when found.

Beyond resolving interactions once they are detected, clever system
designs can automatically select default resolutions. That is, while we
may not be able to anticipate specific interactions or prevent
interactions through isolation, we can anticipate that some interactions
will eventually happen and design the system to handle unknown
interactions gracefully. Automated resolution is usually preferred over
manual resolution if possible, as resolving every interaction
individually is tedious and we may not detect an interaction in time
before some harm is done.

In several domains, automatic \emph{domain-specific default resolution}
mechanisms have been very successful~\cite{bocovich2014variable,JZ:TSE98,Zave2015-zs}. For example, in self-driving cars,
multiple components (cruise control, emergency braking, map) may provide
possibly conflicting suggestions for the target speed (i.e., interaction
through a specific target variable). Anticipating these conflicts, the
system can be designed to resolve these conflicts by \emph{always}
picking the lowest (safest) suggested speed~\cite{bocovich2014variable}.
In Android, it is impossible to foresee all app interactions, but the
system is designed to ask the user when multiple apps can respond to a
request, such as playing a video in Figure~\ref{fig:android}---again a default resolution
strategy is baked into the system design. It is important to note that
these strategies usually need to be designed for a specific problem,
which is their strength and weakness at the same time: on the one hand,
the system can resort to resolutions that leverage domain knowledge,
but, on the other, it is difficult to transfer this type of solution to
other domains.

Default resolution strategies are usually designed for a target domain
such that they usually yield a good, best-effort result, without having
to foresee all possible interactions. Even if the result may not be the
perfect answer, because the component specifications are truly
incompatible (e.g., overlapping preconditions, conflicting goals), a
resolution strategy is typically designed to pick a consistent answer
(e.g., following priorities) that is safe for the task at hand.

The design of many machine-learning systems that are composed of
multiple models seem to already align quite well with this strategy of
resolving interactions at specific communication points: Machine-learned
components are already largely isolated (and wouldn't directly access or
modify each other's internal state) and communicate only through limited
messages. For example, \emph{data fusion} steps to combine outputs of
multiple models are common (e.g., visible explicitly in the Apollo
architecture above). The fusion strategy can either be defined manually
(like picking the lowest speed) or learned with another machine learning
model (often called a meta model)---in both cases, it is usually a
custom solution for the specific task at hand rather than a
general-purpose fusion strategy. For example, a smart thermostat can
learn the behavior of various smart home components and their
interactions to learn to coordinate them better.

Interestingly, in our motivating image captioning scenario, the ranking
component can be seen as a domain-specific fusion mechanism that
combines outputs of object detection and language model. This ranking
component is trained on task-specific data, so it may very well learn
how to cover for inaccuracies in other components and how to resolve
possible interactions. For example, it could learn to not trust
sentences with ``persons'' in it, because it has identified problematic
interactions of how the language model interacts with inaccuracies of
the object detector on those entities. In this sense, the ranking
component is carrying out coordination logic to resolve interactions
that has been learned itself!

We suspect there are many opportunities to think very deliberately about
communication channels and formats to restrict the kind of data exposed
from models and more importantly data fusion steps to define or even
learn default resolutions for interactions.

\hypertarget{different-reasons-for-decomposition}{%
\section{Different Reasons for
Decomposition}\label{different-reasons-for-decomposition}}

In traditional software systems, we need to use a divide and conquer
approach to handle complexity with our limited human cognitive capacity,
and specifications are essential to make this work so that we can focus
on one component at a time. Interestingly, machine learning algorithms
do not share the same capacity limits; they excel at learning models
with millions of parameters, finding patterns in high-dimensional
feature spaces in huge data sets.

The primary reasons for decomposing machine learning problems seems to
be (1) to infuse \emph{structure} and \emph{domain knowledge} into the
solution and (2) to reuse models. By decomposing the image captioning
problem, we can make it explicit that we think object detection and
creating common-sense sentences with those objects are important
ingredients in a solution, e.g., that we only present results generated
by the language model. We can also reuse ML components trained on
different data sets, such as learning the language model on a much
larger dataset of captions without requiring corresponding images. By
reusing ML components trained on other data, we may transfer some form
of domain knowledge that we think will be a useful part of a solution.
However, decomposition is only one of multiple possible design
strategies to encode domain knowledge and reuse, others including
feature engineering, tailoring deep neural network architectures and
inductive biases, transfer learning, and augmenting training data.

The fact that decomposition is a \emph{design option} in machine
learning and \emph{not a necessity} is very visible in the different
submissions for the image captioning competition: some use
decomposition, others learn a single model. Similarly, in research on
self-driving cars, there is a tension between composing many models (as
the Apollo architecture shown above) and learning a
single end-to-end model that
predicts outputs like steering angle directly from raw sensor inputs~\citep[e.g.,][]{bojarski2016end}.

Furthermore, we use machine learning usually to reason about the world
instead of abstract mathematical concepts. However, when reasoning about
humans, vision, or languages, we have to accept that we are working with
real-world concepts that do not compose nicely, concepts we cannot
specify well, and concepts we might not even understand even when a
model has learned them well. Hence it is not surprising that learned
models do not compose nicely either.

\hypertarget{a-call-for-systems-thinking-and-designing-for-interactions}{%
\section{A Call for Systems Thinking and Designing for
Interactions}\label{a-call-for-systems-thinking-and-designing-for-interactions}}

Even though the lack of proper specifications may make machine-learning
models appear special as compared to traditional software systems, there
are many parallels around how to design a system to anticipate
interactions with a healthy dose of system thinking. That is, we need to
focus on \emph{system} design, not just the design of ML model
architectures. Even though the reason for decomposition may be
different, feature interaction thinking is still important, maybe more
than ever.

The key point is to realize that decomposition without perfect
modularity is okay. Decomposition is a best effort approach, but we need
to anticipate interactions, prepare for them as part of the system
design and development process, and make feature interactions a
first-class concern to reason about them when they occur. We need to
embrace design methods of managing interactions, and machine learning
itself may provide a powerful tool in our toolbox for learning
interaction resolution strategies.

At the same time, it is worth exploring what kind of specifications,
however partial, we can provide for machine-learned models. Describing
goals of models or assumptions made in training data selection or
modeling can help to reason at least partially about compositions.
Recent adoption of more structured documentation like model cards and
datasheets can provide inspiration for providing structured and possibly
even machine-readable information about models.

\bibliographystyle{plainnat}
\bibliography{bibl,dblp2_short}

\end{document}